# Stability of radical-functionalized gold surfaces by self-assembly and on-surface chemistry

Tobias Junghoefer,[a],[‡] Ewa Malgorzata Nowik-Boltyk,[a],[‡] J. Alejandro de Sousa,[b,d] Erika Giangrisostomi,[c] Ruslan Ovsyannikov,[c] Thomas Chassé,[a] Jaume Veciana,[b] Marta Mas-Torrent,[b] Concepció Rovira,[b] Núria Crivillers,[b] Maria Benedetta Casu*[a]

[a]Institute of Physical and Theoretical Chemistry, University of Tübingen, 72076 Tübingen, Germany

[b]Institut de Ciència de Materials de Barcelona (ICMAB-CSIC) and Networking Research Center on Bioengineering Biomaterials and Nanomedicine (CIBER-BBN) Campus de la UAB, 08193 Bellaterra, Spain

[c]Helmholtz-Zentrum Berlin für Materialien und Energie (HZB), 12489 Berlin, Germany.

[d]Laboratorio de Electroquímica, Departamento de Química, Facultad de Ciencias, Universidad de los Andes, 5101 Mérida, Venezuela

**Abstract:** We have investigated the radical functionalization of gold surfaces with a derivative of the perchlorotriphenylmethyl (PTM) radical, using two methods: by chemisorption from the radical solution and by on surface chemical derivation from a precursor. We have investigated the obtained self-assembled monolayers by photon-energy dependent X-ray photoelectron spectroscopy. Our results show that the molecules were successfully anchored on the surfaces. We have used a robust method that can be applied to a variety of materials, to assess the stability of the functionalized interface. The monolayers are characterized by air and beam stability unprecedented for films of organic radicals. Over very long beam exposure we observed a dynamic nature of the radical-Au complex. The results clearly indicate that (mono)layers of PTM; radical derivatives have the necessary stability to stand device applications.

INTRODUCTION

Molecular systems are materials that intersect with many different promising fields such as organic/molecular spintronics, electronics, and organic magnetism.[1-8] In this framework, organic radicals are exceptionally promising in various fields, and the research on radical thin films and interfaces has recently flourished, due to their potential use in applications from quantum computing to organic electronics and spintronics[8-13].

We have recently demonstrated that a Blatter radical derivative is a potential quantum bit and we attached it to copper contacts to investigate the influence of a substrate on the radical magnetic moment.[9] Our work indicated the need for identifying strategies in order to attach the radical to the surface preserving its magnetic moment at the interface, by using different methods ranging from evaporation to preparation in wet environment. However, the radical functionalization of a substrate is eased choosing a specific chemical group that has a high chemical affinity for the selected substrate. Usually thiols and disulfides are chosen to covalently modify gold surfaces, including gold nanoparticles, with organic radicals by adsorption from solution. More recently alkyne terminated derivatives start to play a role. Nitroxides (TEMPO),[14-17] nitronyl nitroxides[18-20] and tripheylmethyl[21-23] radicals have been successfully employed to prepare such paramagnetic hybrid materials. In this work, we capitalize our knowledge of radical thin films and interfaces studying the functionalization of gold surfaces with derivatives of the perchlorotriphenylmethyl (PTM) radical. PTM is a very persistent and stable radical that shows long coherence time at room temperature, being a strong potential candidate for quantum technologies.[24] Previously, self-assembled monolayers (SAMs) of PTM on gold substrates have been investigated to study their transport properties.[21-23, 25] The radical character of the layers was proved by several techniques (UV-vis, cyclic voltammetry, EPR, NEXAFS and UPS), however, a careful and in-depth characterization of the stability of these radical SAMs was not carried out so far. Such stability is a necessary precondition to use radical-based SAMs for any practical application. Here, we used a ferrocene functionalized PTM derivative with an alkyne termination (Figure 1) that covalently attaches to the gold substrate spontaneously.[26-31] The ferrocene functionalization makes the molecules interesting for current rectification, as seen in SAMs incorporating ferrocene acting as redox-active moiety.[32-34] We investigated also the formation of radical self-assembled monolayers (SAMs) obtained by using on-surface chemistry.

Our investigations were performed using X-ray photoelectron spectroscopy (XPS). While XPS is a well-established technique to investigate the electronic structure of materials, this is not the only aspect that can be examined.[35] Because of its high sensitivity, it is also possible to quantitatively calculate the stoichiometry of the investigated systems. Further aspects can be explored: it is very sensitive to the chemical environment of the elements, allowing revealing the occurring chemical bonds and the charge transfer from/to surfaces. It is possible to gain information on film stability (e. g., under beam or air exposure) and on post-growth phenomena. It is extremely well suited to investigate radical thin films (including their radical character) when evaporated by using controlled conditions.[36] We proved in our previous work that XPS in combination with a careful and robust best fit procedure allows investigating the radical character, being the results in perfect agreement with electron paramagnetic resonant (EPR) measurements.[9, 36-41] EPR is the technique typically used for radical characterization. However, its use for films is limited 1) by the fact that is an ex-situ technique. Radical thin films might not be stable enough outside the ultra-high vacuum environment where they are deposited or obtained by on surface reaction.[42-44] 2) By the choice of the substrate that might contribute to the EPR signal.[41] 3) By the substrate dimensions that are often over-dimensioned for standard spectrometers. 4) Standard EPR Spectrometer do not have the necessary sensitivity to measure (sub)monolayers. Conversely, XPS has a high sensitivity further beyond many other conventional chemical techniques, as it can detect less than $10^{13}$ atoms,[45] allowing investigations in the monolayer and submonolayer regime without requiring advanced "state of the art" spectrometer, as it is the case for EPR, but a standard, commercially available, monochromatized laboratory XPS station is sufficient.

In this work, we investigate the chemistry of the SAMs/gold interface, demonstrating that the SAMs were successfully attached to the substrate, using also on-surface chemistry. We also show that it is possible to identify the spectroscopic lines associated to a radical character versus its diamagnetic counterpart. The work focuses on the SAM stability, under X-ray and air exposure, using a method that can be applied to any material to explore any kind of stability issues, such as, gas exposure, humidity, aging, temperature that are of paramount importance for technological applications.

EXPERIMENTAL SECTION

SAM 1 and SAM2 were prepared following the protocol thoroughly described in Ref. [34]. SAM4 was grown following a two steps reaction: 1) SAM1 was immersed in a 2 mM solution of $Bu_4NOH$/THF(freshly distilled) under argon atmosphere. The solution was left with a gentle stirring for 8 h at room temperature in dark. Then, the substrates were removed from the flask and thoroughly rinsed with THF (distilled). 2) Immediately afterwards, the substrates were immersed in a 4mM p-chloranil/THF(distilled) solution under argon atmosphere. The solution was left 12 h at room temperature in dark. Finally, the substrates were removed from the flask, thoroughly rinsed with THF(distilled) and dried with a nitrogen stream. Coverage and radical formation were checked with cyclic voltammetry.

The XPS Ultra High Vacuum (UHV) system ($2 \times 10^{-10}$ mbar base pressure) was equipped with a monochromatic Al K$\alpha$ source (SPECS Focus 500) and a SPECS Phoibos 150 hemispherical electron analyzer. Survey spectra were measured at 50 eV pass energy and individual core level spectra at 20 eV pass energy. Both were subsequently calibrated to the Au 4f signal at 84 eV. To minimize potential radiation damage, freshly prepared films were measured, and radiation exposure was minimized unless differently stated in the text (i.e., stability measurements). For measurements probing air stability, beam exposure was further limited after air exposure to attribute the observed changes exclusively to the degradation by air exposure.

Photon-energy dependent XPS measurements were performed at the third-generation synchrotron radiation source BESSY II (Berlin, Germany) at the Low-Dose PES end station installed at the PM4 beamline (E/ΔE = 6000 at 400 eV). They were carried out in multibunch hybrid mode with a SCIENTA ArTOF electron energy analyzer (ring current in top up mode = 300 mA).

SAM2

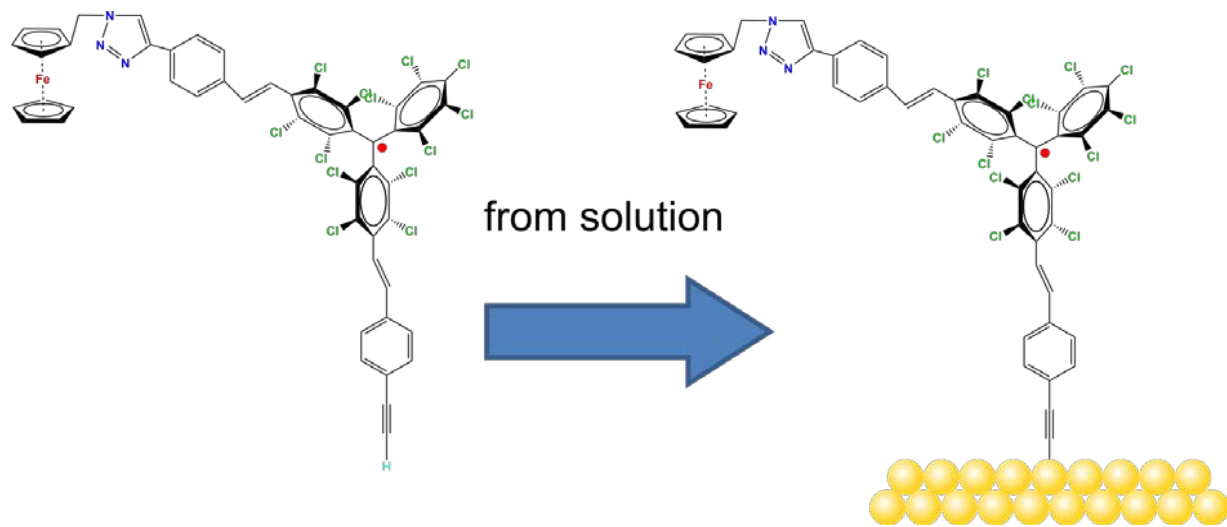

SAM1 and SAM4

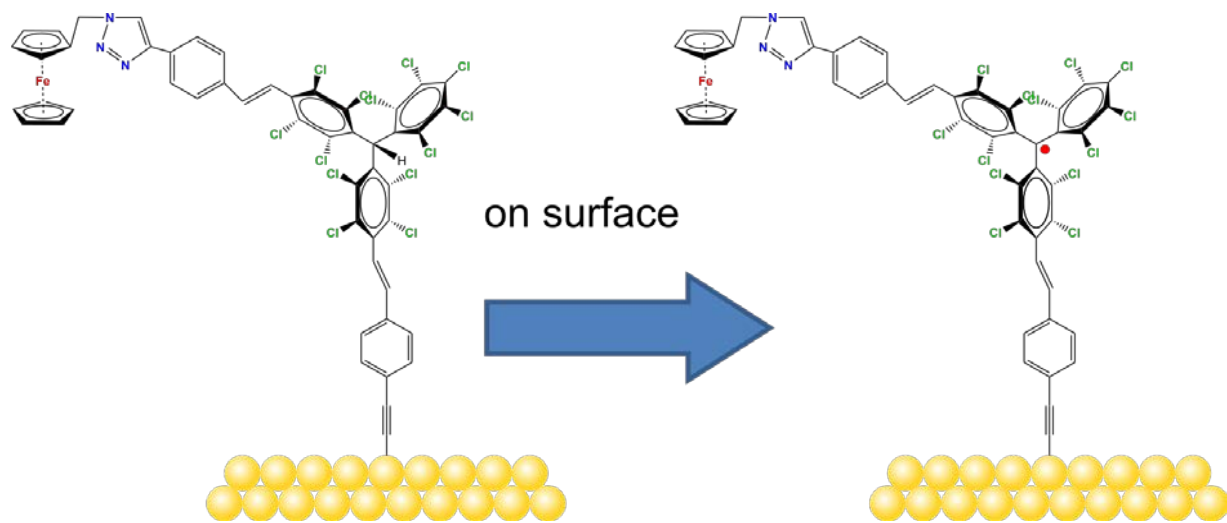

Figure 1. Molecular structure of the radicals, as indicated, and schematic sketch of the different SAM preparations.

RESULTS AND DISCUSSION

We examined two different layer preparations using the PTM radical derivative (SAM2 and SAM4) and we compared them with those obtained depositing the diamagnetic counterpart,

SAM1 (Figure 1). The PTM radical and the diamagnetic derivative shown in Figure 1 were synthesized as previously reported:[34] SAM2 is obtained depositing the radical on the gold substrate from its solution. SAM4, on the contrary, is obtained by first depositing the analogous diamagnetic molecules on gold and following a two-step synthesis (i.e., anion generation and oxidation), thus, the PTM radical is formed on surface.[46]

Figure 2 shows the SAM2 XPS spectra of the important core levels (for a survey, and the stoichiometric analysis, see Figure S1 and Tables S1 and S2 in the electronic supplementary information (ESI)). The spectra are characterized by the predominance of the gold signals in agreement with the deposition of a monolayer. Apart from a carbon concentration that slightly exceeds the theoretical values, usual in samples prepared ex-situ with wet-environment techniques, the films are remarkably clean, and no significant amounts of contaminants are visible. In XPS, the integrated area of the main lines corresponding to photoelectrons emitted from a given element, together with their satellites, is proportional to the concentration of that same element in the investigated system.[35, 47, 48] In highly resolved XPS spectra, the rich fine structure allows fitting the lines including contributions for different atomic sites of the same element that due to a different chemical environment are expected to show differences in their binding energies.[35, 47, 48]

The film stoichiometry agrees with the expected values, confirming that the radical derivative was indeed attached to the gold substrate. The C 1s spectroscopic line is characterized by a main peak at around 284.5 eV and a picked feature at around 286 eV. The C 1s intensity is due to photoelectrons emitted from the carbon atoms. The contributions mirror several different chemical environments. In fact, carbon atoms are not only bound to other carbon atoms, but to hydrogen, nitrogen, and chlorine atoms. Each different environment leads to a slightly different binding energy that can be identified by using a best fit procedure (Figure 2a).[34, 37, 49] The fitting procedure in XPS is driven by specific and detailed chemical and physical arguments, and not by a mere mathematical approach. The curves are described by a Voigt profile, i.e., a convolution of a Gaussian and a Lorenzian profile. This is because different contributions influence the line shape of the XPS main features: intrinsic lifetime broadening, vibronic and inhomogeneous broadening lead to a Lorenzian profile, while experimental contributions have a Gaussian profile. The lifetime of the core hole is determined basically by the Heisenberg uncertainty principle and consequently the intrinsic peak width is, too. For example, the Lorenzian width for the C 1s orbital is around 80 meV, and for the N 1s orbital is around 100 meV in organic materials.[50] The experimental setup gives a contribution assumed to have a Gaussian lineshape due to the resolution of the analyser, the non-perfect monochromaticity of the X-ray, and inhomogeneities of different nature. The fit that we use is based on the procedure adopted for closed-shell molecules.[14] The final fit is the result of several self-consistent interactions of sequential fits done considering all physical and chemical information and adding more constraints at each interaction, with the goal to keep the parameter dependency very low (dependency values of the last fits in this work were very close to zero). The constraints in our fit are based on the element concentration, the binding energy constraints must adhere to electronegativity and known values in literature, we use the published or measured core-hole lifetimes for each element. The fit procedure must systematically hold for all samples of a specific system, prepared, and measured under the same conditions. Our procedure revealed to be extremely robust giving results in very nice agreement with EPR and ab-initio calculations, both for open-shell and closed-shell systems, as well.[9, 36-41, 51-55] To reach this result, we work on sets of samples that are large enough to be statistically significant. In this way, we can also identify the samples that do not correspond to the expected stoichiometry.[37, 39, 56, 57]

In the spectra, we observe the presence of shake-up satellite intensities (Figure 1). As a result of the core-hole formation the symmetry is reduced, and a larger number of non-equivalent carbon atoms should be considered.[58, 59] The ionization at different carbon sites may give different contributions to the shake-up spectra. The $S_1$ satellite can be related to the first

HOMO-LUMO shake-up.[60] Its energy position with respect to the main line is lower than the optical gap, a typical effect in the HOMO-LUMO shake up satellites of polyaromatic molecules caused by the enhanced screening of the core-hole due to its delocalization.[60-63] A large number of satellite features is expected upon a photoemission event. However, their assignment is very complicated, especially for large molecules because not completely described by theoretical models. Such detailed description is outside the goal of this work; therefore, we have identified most of the higher binding energy satellite intensities under a unique component, $S_2$. This component is correlated with the C-Cl feature from a stoichiometric point of view. This assignment is further corroborated by the fact that the C-Cl feature and $S_2$ change simultaneously depending on the photon energy, as it can be easily seen in Figures 2 and 3.

The intensities of the several contributions agree with the expected stoichiometry, confirming once more that the SAM2 carbon line correspond to the radical derivative (Table S2 in the ESI). The Cl 2p, N 1s and Fe 2p core level spectra are also shown. Their features confirm the presence of an intact molecule: Cl 2p core level lines show the typical doublet feature (spin-orbit splitting = 1.6 eV as in the literature[64]), the N 1s spectrum (Figure 2b) is characterized by contributions due to photoelectrons emitted from three different chemical environments, confirming the intactness of the triazole derivative (Figures 2c and 2d). The signal of the Fe 2p shows the expected doublet (spin-orbit splitting = 12.8 eV concomitant with the values in the literature[64, 65]), and the noteworthy absence of further intensities indicating that the signal is due to electrons emitted from iron atoms in the +2 oxidation state, as it is the case for ferrocene.[65-67] Note that in monolayers of clean ferrocene, the Fe 2p spectrum does not show any satellite intensities.[65-69] In fact, this intensity depends on the ligands and it varies with their electronegativity.[70] Additionally, in the case of monolayers on metal substrates, the image-charge formed at the interface [71] may further screen the satellite intensities. The XPS intensities and line shapes indicate that the radical was attached to the surface preserving the expected stoichiometry. Thus, we can confidently infer that the synthesis and the preparation of SAM2 were successful.

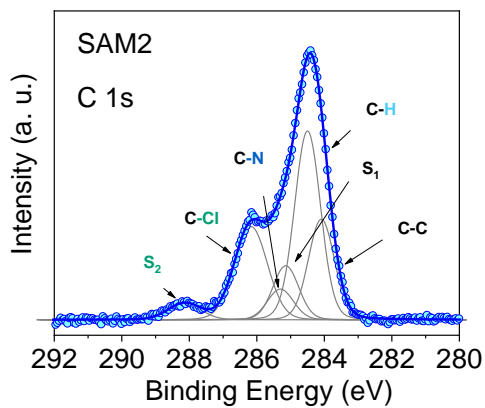
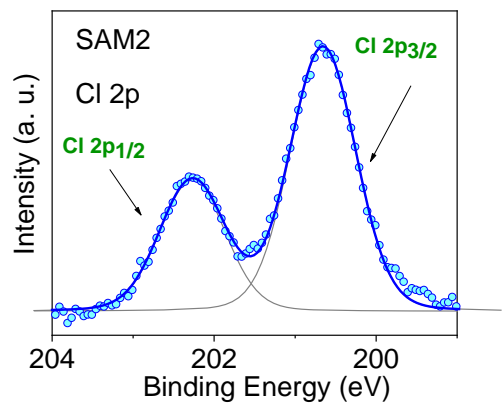

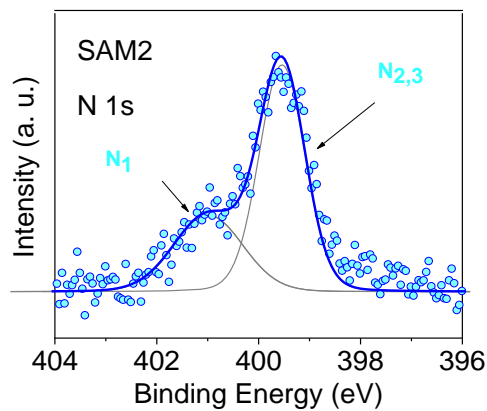
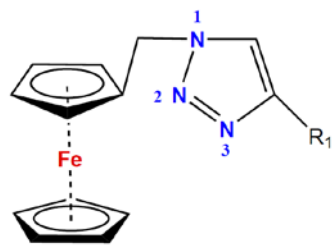

a)　　　　　　　　　　　　　　　　　　　b)

c)　　　　　　　　　　　　　　　　　　　d)

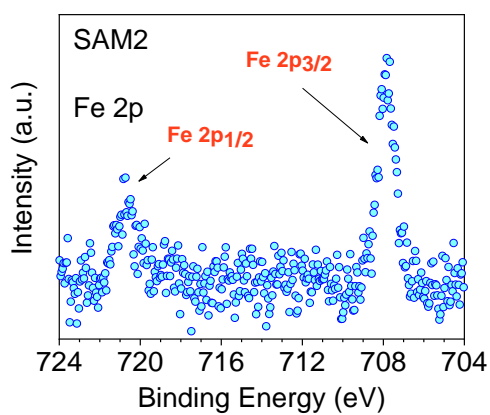

e)

Figure 2. SAM2. a) C 1s, b) Cl 2p, c) N 1s (together with their best fit) and e) Fe 2p XPS spectra (photon energy: 1486.6 eV). In d) the chemical environment of the triazole derivative is shown in detail.

To support this conclusion and explore the use of XPS to identify the PTM radical, we investigated SAM1, i.e., the SAM obtained from the diamagnetic counterpart of the PTM radical derivative (Figure 1).

The essential core level spectra are shown in Figure 3 (for a survey, and the stoichiometric analysis, see Figure S2 and Tables S3 and S4 in the ESI). In our discussion, we focus on the C 1s core level spectroscopy line. This is the line that is directly correlated with the radical character (see Figure 1) because the unpaired electron mainly resides in the central radical carbon atom of the perchlorinatedtriphenylmethyl unit. The stoichiometry for SAM2 and SAM1 is different. In SAM1 the central methyl carbon atom of the PTM is bound to hydrogen. Therefore, we expect a different C 1s line broadening with respect to the radical spectra. Indeed, we observe a larger line for SAM1 (Full Width at Half Maximum (FWHM) =1.8 eV versus 1.4 eV for SAM2, under the same experimental conditions). This difference is mirrored by a larger gaussian width required in the fit procedure (see Table S4 in the ESI). We also observe a different binding energy. SAM1 C1s main line is at higher binding energy than SAM2 main line. This indicates that the core-hole created upon photoemission is more efficiently screened in SAM2 than in the diamagnetic molecule. This can be explained considering the donor-acceptor character of SAM2[34] where the simultaneous presence of the radical and the azidomethyl ferrocene unit stands for faster charge delocalization of the core-hole. These differences in the C 1s main line, binding energy and broadening between SAM1 and SAM2 allows using XPS to identify the radical character of the SAMs.

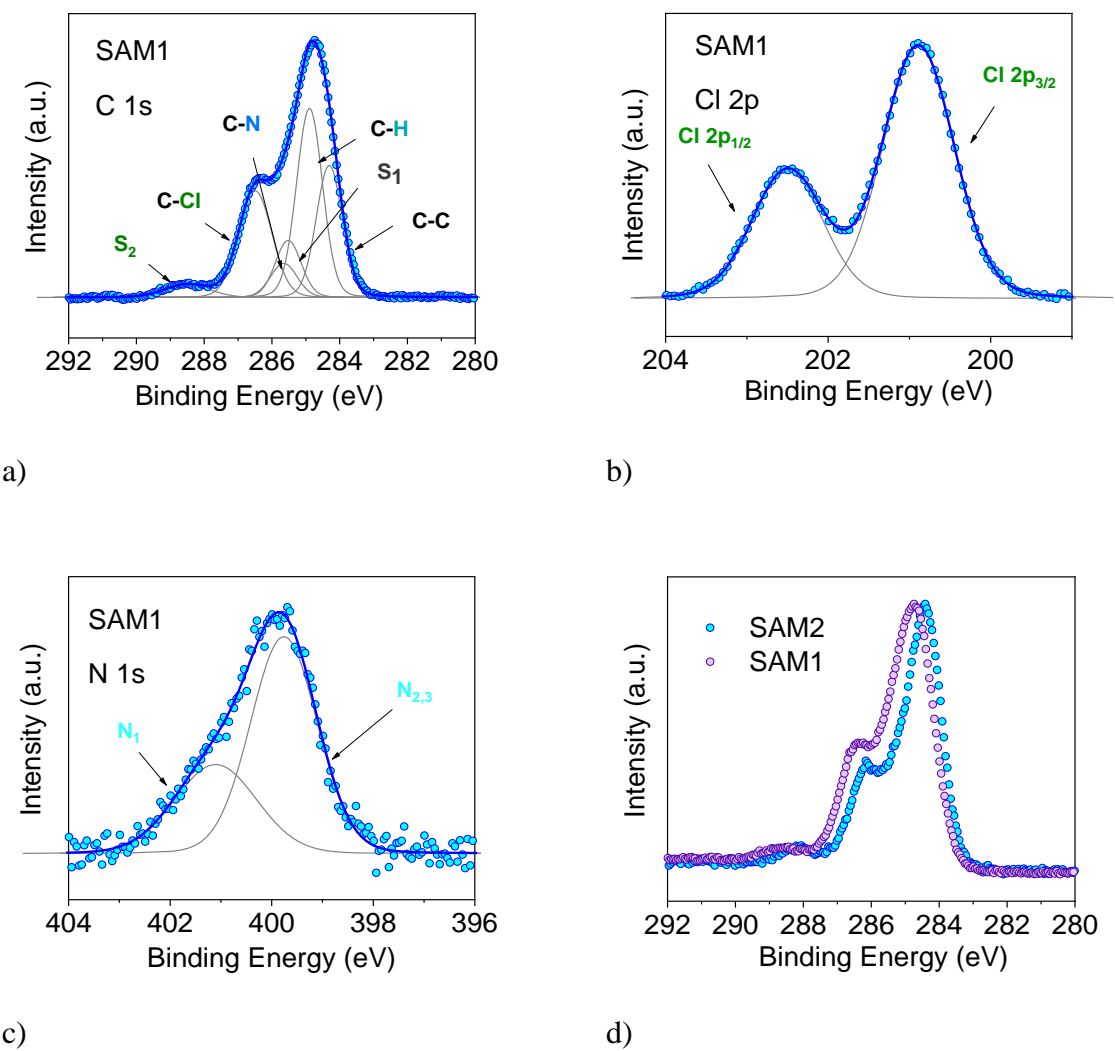

Figure 3. SAM1. a) C 1s, b) Cl 2p, and c) N 1s (together with their best XPS spectra (photon energy: 1486.6 eV). d) Comparison of the C 1s XPS line of SAM1 (diamagnetic) and SAM2 (radical). Intensities are normalized to the peak maximum to allow comparison.

In a XPS experiment it is possible to probe different sampling depths:[72] changing the photon energy, the materials emit electrons with different kinetic energy that is equivalent to emit photoelectrons with different inelastic mean free path ($\lambda$). Thus, we have performed a photon-energy-dependent experiment on SAM1, SAM2 and SAM4 using 460 and 640 eV photon energy, respectively. This corresponds to vary $\lambda$ between 0.17 and 0.28 nm[73, 74] (Figure 4 and Figure S3). The experiment at 460 eV is very surface-sensitive (note that both experiments, at 460 and 640 eV are very surface-sensitive with respect to the measurements so far discussed, which were performed at 1486.6 eV). We observe that varying the photon energy the relative intensities of the main line and the line due to photoelectrons emitted from carbon atoms bound to the electronegative nitrogen and chlorine atoms change: the feature at higher binding energy has higher intensity at 640 eV. What is also important is that these changes are accompanied by changes in the $S_2$ satellite, indicating, as mentioned, that these two components are strongly correlated, corroborating our fit assignments. This change in the intensity depends on the photon energy and, thus, on the inelastic mean free path and it is due to the surface core level shift effect,[75-78] i.e., the difference of the core level photoemission from a surface atom/molecule with respect to a bulk atom/molecule.[76, 78, 79] This effect is visible in organic thin films when the molecules are not planar, and carry electronegative atoms.[49, 80, 81] In fact, electronegative atoms shift the electronic cloud, causing a different screening of the core-hole created upon photoemission. However, this screening is different when occurs at different depths where structural differences are significant, for example, in case of upright versus flat lying molecules.[80] In the present case, the C-Cl components is stronger at 640 eV when the experiment is less surface-sensitive. We can infer a structural information from this dependence: the XPS results indicate that the PTM radical is closer to the substrate with respect to the azidomethyl-ferrocene unit, (as sketched in Figure 1) therefore its contribution is stronger when $\lambda$ is longer. For the photon energy 1486.6 eV, $\lambda$ is comparable with the dimension of the molecule ($\lambda = 0.81$ nm[73]), in which case the stoichiometry information plays the major role against the structural information, as seen in closed-shell systems like phthalocyanines.[82]

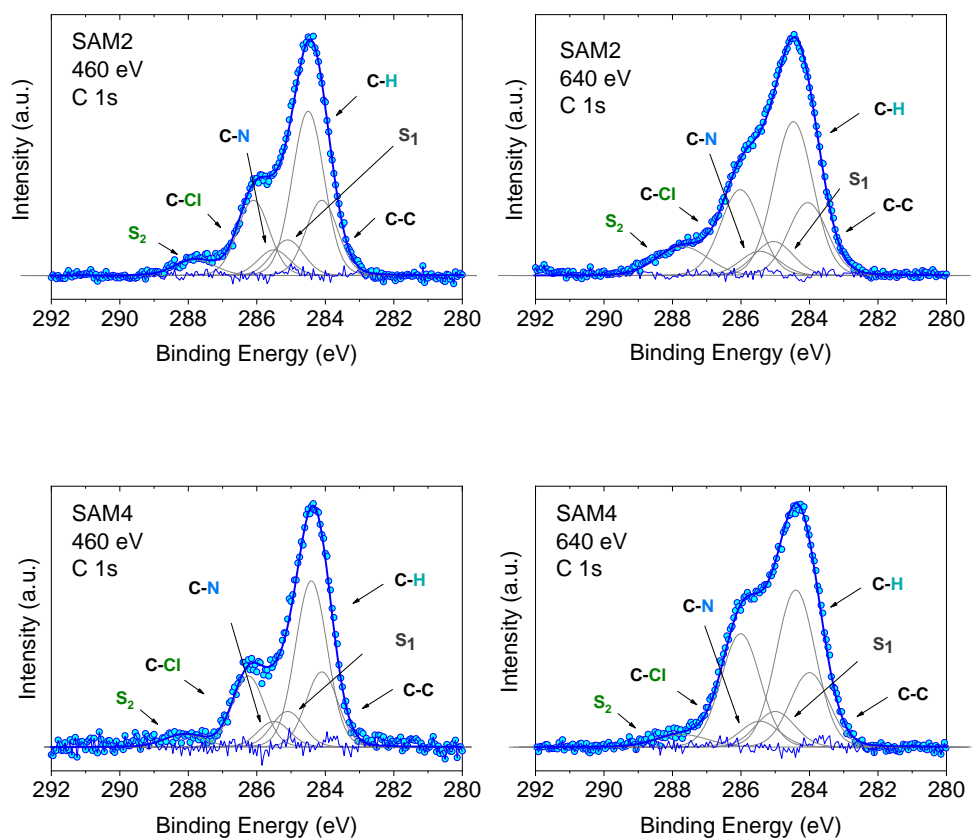

Figure 4. C 1s core level spectra at 460 and 640 eV, as indicated, together with their best fit and residua. Upper panel: SAM2. Lower panel: SAM4. Intensities are normalized to the peak maximum to allow comparison. For the curve fits see the ESI.

Using the above results as a reference, we investigated SAM4 (Figure 4, lower panel). This monolayer has the same theoretical stoichiometry as SAM2, but it has been obtained via on-surface radical formation from the diamagnetic molecule. We focused once more on the C 1s core level spectra. First, from the point of view of the stoichiometry, as previously done for SAM2 and SAM1. We observe that the C 1s line shape had the same features as in the SAM2 core spectra. Also, in this case, we observe the same photon energy dependence at 460 and 640 eV, hinting at a similar structural adjustment of the molecule units with respect to the substrate.

What is most important is that the FWHM of the C 1s line is narrower than in the case of the diamagnetic molecule, i.e., SAM4 has a narrower main line than SAM1 (see Figure S3). Following our above discussion, this effect indicates a radical character of the film. Since the radical generation was performed on surface, this result hints at and support the successful on-surface preparation of the radical. A fit procedure back these observations: the same best fit procedure leads to same intensities and binding energies for the C 1s contributions of the spectra of SAM2 and SAM4 (Figure 4 and Tables S5,S6, S7 and S8 in the ESI). Cyclic voltammetry experiments support the radical character of the layers, too (see Figure S4 in the ESI). The redox peaks corresponding to PTM radical ↔ PTM anion and ferrocene ↔ ferrocenium redox process are clearly observed.

A change in photon energy as performed in the present XPS experiments also implies a change in the C 1s cross section increasing the complexity of the screening effects. Looking at the fit results, we note that the $S_1$ intensity decreases with increasing the photon energy while the intensities of the $S_2$ satellite shows the opposite behaviour (Tables S5-S8 in the ESI). This gives a hint about the fact that the S1 intensity is related to the dipole excitation of a core electron to the lowest unoccupied molecular orbital (LUMO) accompanied by the monopole ionization of the valence electron: this shake up contribution is near to the ionization threshold region and decreases with the increase in energy,[83, 84] as observed in our fits.

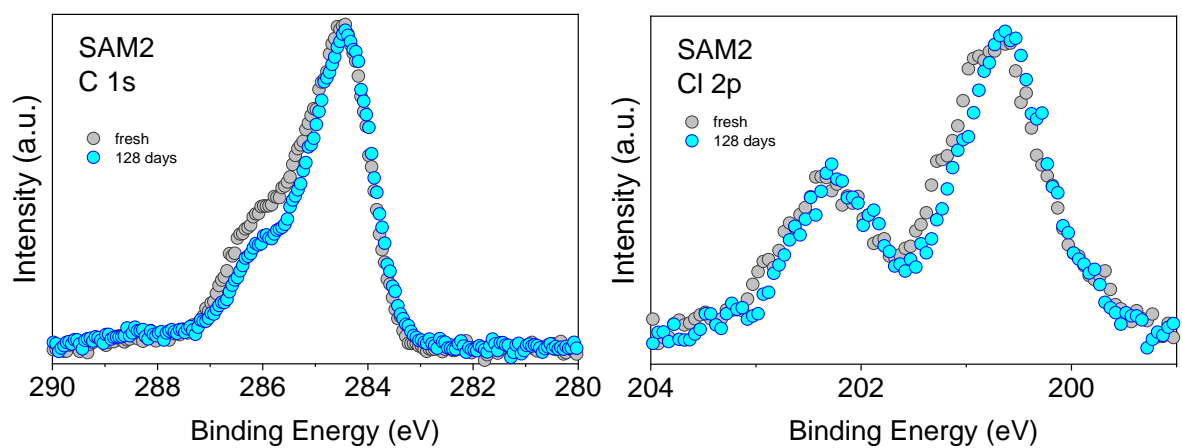

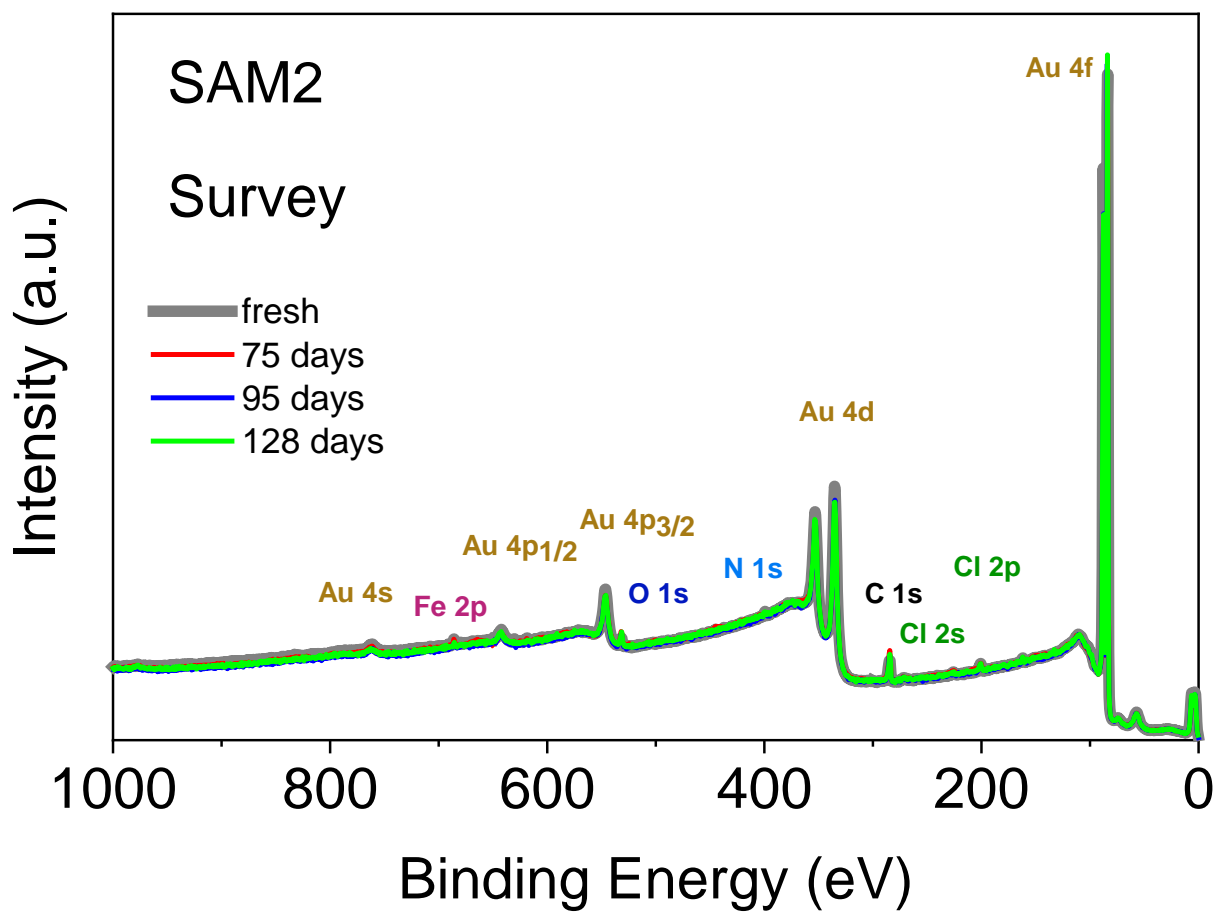

Figure 5. Upper panel. (left) C 1s and (right) Cl 2p core level spectra for a freshly prepared monolayer and after 128 days exposed to air and kept in darkness, as indicated (photon energy: 1486.6 eV). Intensities are normalized to the peak maximum to allow comparison. Lower panel. SAM2 Survey XPS spectra under air exposure, as indicated (photon energy: 1486.6 eV).

An important aspect that we intend to address here is the stability of the monolayer in real environment. While the PTM radical is known to be chemically stable both in solution and in powder if visible light is avoided, there is no report on the chemical and structural stability of its films where single radical molecules are exposed to air. To tackle this issue, we kept SAM2 monolayers under air in darkness and measured them again 128 days later, always minimizing X-rays exposure during measurements. The results are shown in Figure 5. The C 1s core level spectrum comparison between the fresh monolayer and the "aged" monolayer shows a small difference in the relative intensity of the main feature with respect to the feature at higher binding energy, while the Cl 2p spectra do not show major differences. Post growth phenomena, such as desorption and ripening are expected and well-known in case of organic molecules, and expected also in radical films, especially for those systems having low vapor pressure at room temperature and physisorbed on surfaces.[36, 37, 40] To investigate the origin of the difference in the C 1s core level spectra we performed a best fit analysis, following two hypotheses. In one case, we performed the fit considering that PTM might switch to the perchlorophenylfluorenyl radical (PPF) (Figure S5, Table S9 in the ESI). This is a known derivative of the PTM radical generated both by heating over 300 °C[85] or by photoirradiation.[86] In the second case, we considered that the stoichiometry of the monolayer stays unchanged but the carbon intensity increases due the adsorption of carbon impurities from the environment (Figure S6 and Table S10 in the ESI). Both fits are plausible. A closer inspection of the survey spectra helps to interpret the results (Figure 5, lower panel). Initially, the gold signal is stronger, i.e., its intensity decreases with time. Simultaneously the carbon signal increase, while the chlorine signals does not change. From the stoichiometric analysis of the spectra, we found that in the fresh monolayer the carbon to gold ratio (C/Au) and the chlorine to gold ratio (Cl/Au) are 0.37 and 0.04, respectively. After 128 days, they are 0.40 and 0.04, respectively. That clearly indicates that the chlorine content does not diminish and that the phenomenon playing the major role is carbon adsorption. This means that not only the PTM radical is chemically stable, but also its monolayers are stable under prolonged air exposure. A result of great significance because it fully supports the use in devices of the PTM radical and its derivatives grafted on surfaces.

We also studied the stability of SAM2 against X-ray. As previously done, we focus our discussion on the PTM radical analysing the C 1s and the Cl 2p core level spectra (Figure 6). We could observe first small changes in the spectroscopic lines after 18 hours X-ray exposure, a 0.1 eV shift of the binding energy towards higher values and a difference in the satellite intensities. The fit analysis performed on the C 1s line confirms that these are no significant stoichiometric changes (Figure S7 and Tables S11 and S12 in the ESI). We crosschecked this finding also using synchrotron radiation and monitoring the film in real-time over around 8 hours (Figure 6, lower panel, photon energy: 640 eV, flux: 1 x $10^9$ - 1 x $10^{10}$ photons/s). No changes were detected.

To understand what happens under very long X-ray exposure, we exposed the films to X-rays for 52 hours and we looked at the effects (Figure 6d). After such a long exposure, the gold signal is more intense, while the C 1s and Cl 2p lines show no decrease in the intensity. This indicates that the gold substrate is more exposed with time. Usually this result hints at changes in the film morphology due to post-growth phenomena, such as desorption, dewetting or Ostwald ripening, that lead to the coalescence of small islands into big islands leaving the substrate surface free. The result indicates, also in this case, some degree of dynamics, suggesting a change in the layer morphology. These experimental observations seem puzzling in the case of a strong adsorbate-substrate chemical bond. To help in understanding this phenomenon, we can look at one of the most investigated SAM systems: the thiolates on gold. Investigations of the thiolates-Au surfaces have demonstrated a clear dynamic nature of this surface, where the mobility of the adsorbate-Au complex plays an important role, both on flat surfaces as well as on nanoparticles, upon mild annealing and even at room temperature.[87] The mobility is explained in terms of presence of defects on the gold surfaces.[87-90] At defect sites, the interaction between the single gold atom and the covalently attached molecule is stronger than the interaction with the environment (gold atoms and surrounding molecules, respectively) causing the motion of the complete adsorbate+Au assemble on the surface, giving rise to ripening, and even to desorption. This is a very general mechanism of surface diffusion occurring when the adsorbate is strongly bound to coinage metals as gold.[87, 89] The behaviour of the PTM-based SAMs on gold and the consequent XPS observed during long beam exposure would hint at that fact that such mechanism also occurs in the present case, favoured or induced by the prolonged X-ray exposure.

CONCLUSIONS

Once more, XPS has proved to be a very powerful tool to investigate radical films and radical/metal interfaces, uncovering phenomena not yet known. Furthermore, our XPS method to assess the stability of radical/inorganic interfaces can be applied to any system. In this work, we have investigated the stability of chemically functionalized gold surfaces with a PTM radical, either by preparing the self-assembled monolayers directly from the radical solution, or alternately, by chemical means obtaining the radical on surface from its diamagnetic precursor. While the chemical stability of the PTM radical was well-known, PTM is considered an inert radical, here we show that the radical monolayers have unprecedented stability under ambient conditions and aggressive X-ray exposure. Extremely prolonged X-ray exposure indicates a dynamic nature of the radical-Au complex, analogously to the case of thiolates-Au surfaces. To our knowledge, this phenomenon was not yet reported for this class of adsorbate-Au systems. Therefore, further investigations, including annealing experiments and theoretical modelling, are necessary to deepen the dynamical aspects of this surface. We cannot exclude that similar phenomena might occur at room temperature also upon long air exposure, with reaction time of weeks, as seen for thiolates-Au nanoparticles.[87] Although further investigations on the long-term aging pattern of the PTM radical-based layers also depending on different parameters, such as temperature and visible light, are necessary, our results point out that carbon absorption from ambient plays the major role when the monolayer is exposed to air for a long time.

The PTM radical and its derivatives form monolayers that have unprecedented stability properties, confirming that these systems are suitable candidates for market-oriented applications.

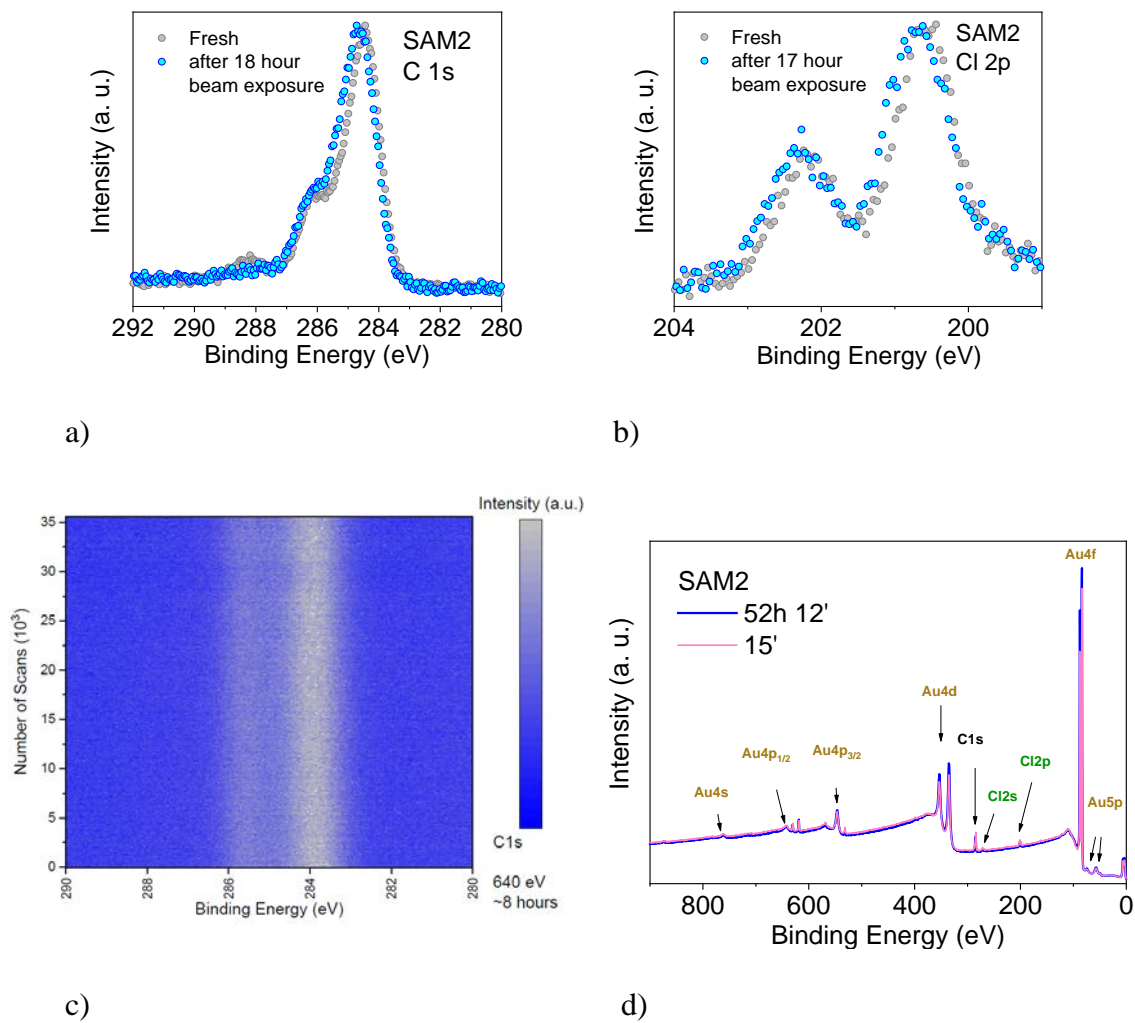

Figure 6. Upper panel: a) C 1s and b) Cl 2p core level spectra of a freshly prepared monolayer and after 18 hours X-ray exposure, as indicated (photon energy: 1486.6 eV). Intensities are normalized to the peak maximum to allow comparison. Lower panel: c) Time-dependent C1s core level signal. Color scale: Blue represents the background signal; white the initial peak intensity. (photon energy: 640 eV). d) Survey XPS spectra of a freshly prepared monolayer and after 52 hours X-ray exposure, as indicated (photon energy: 1486.6 eV).

ELECTRONIC SUPPLEMENTARY INFORMATION (ESI)

Survey spectra of SAM2A and fit results for the photoemission lines in the SAM2A C 1s spectra. SAM1 Survey, stoichiometric analysis, and fit results for the photoemission lines in the SAM1 C 1s spectra. C 1s core level spectra at 460 and 640 eV. Fit results for SAM2 and SAM4 at 460 eV. Fit results for SAM2 and SAM4 at 640 eV. Electrochemical measurements. Stability under air exposure. Stability under beam exposure.

AUTHOR INFORMATION

Corresponding Author

*benedetta.casu@uni-tuebingen.de

Author Contributions

‡These authors contributed equally.

ACKNOWLEDGMENTS

The authors would like to thank Helmholtz-Zentrum Berlin (HZB) for providing beamtime at BESSY II (Berlin, Germany), Hilmar Adler, Elke Nadler, and Sergio Naselli for technical support. J. A. de S. is enrolled in the Materials Science PhD program of UAB. J.A. de S. thanks the Spanish Ministry for the FPI fellowship. This work was funded by the Spanish Ministry project FANCYCTQ2016-80030-R, the Generalitat de Catalunya (2017SGR918) and the Spanish Ministry of Economy and Competitiveness, through the "Severo Ochoa" Programme for Centers of Excellence in R&D (SEV-2015-0496), the CSIC with the i-Link+ 2018 (Ref. LINKA20128) and CIBERBBN. J. A. de S. is enrolled in the Materials Science PhD program of UAB. J.A. de S. thanks the Spanish Ministry for the FPI fellowship. Financial support from HZB and German Research Foundation (DFG) under the contract CA852/11-1 is gratefully acknowledged.